\documentclass[%
 reprint,
 amsmath,amssymb,
aps,
]{revtex4-2}
\usepackage{xcolor}\usepackage{amsmath}
\usepackage{slashed}
\usepackage{hyperref}
\usepackage{graphicx}
\usepackage{dcolumn}
\usepackage{bm}
\usepackage{multirow}

\newcommand{\blu}[1]{\textcolor{blue}{#1}}

\def\n			{\ensuremath{\nonumber\\ }}
\def\no			{\ensuremath{\nonumber\\ &&}}
\def\nou		{\ensuremath{\nonumber\\ &=&}}

\begin{document}

\preprint{APS/123-QED}

\title{Microscopic parametrization of the near threshold oscillations\\ of the nucleon time-like effective electromagnetic form factors}%

\author{Francesco Rosini}
  \email{francesco.rosini@pg.infn.it}
\author{Simone Pacetti}%
\affiliation{%
Dipartimento di Fisica e Geologia, INFN Sezione di Perugia, 06123 Perugia, Italy
}%
\author{Olga Shekhovtsova}
\affiliation{%
INFN Sezione di Perugia, 06123 Perugia, Italy
}%
\affiliation{%
National Science Centre, Kharkov Institute of Physics and Technology, Akademicheskaya, Ukraine
}%

\author{Egle Tomasi-Gustafsson}
\affiliation{%
DPhN, IRFU, CEA, Universit\'e Paris-Saclay, 91191 Gif-sur-Yvette Cedex, France  
}%

\begin{abstract}
We present an analysis of the recent near threshold BESIII data for the nucleon time-like effective form factors. The damped oscillation emerging from the subtraction of the dipole formula is treated in non-perturbative-QCD, making use of the light cone distribution amplitudes expansion. Non-perturbative effects are accounted for by considering $Q^2$-dependent coefficients in such expansions, whose free parameters are determined by fitting to the proton and neutron data. 
Possible implications and future analysis have been discussed.
\end{abstract}

\maketitle

%
%
\section{\label{sec:intro}
Introduction}
The theoretical impossibility of describing the nucleon internal structure in terms of strongly interacting quarks and gluons, which are the fundamental fields of quantum chromodynamics, enhances the electromagnetic form factors (EMFFs) to the role of unique and privileged tools to unravel the dynamics underlying the electromagnetic interaction of nucleons. They provide the most effective description of the mechanisms that determine and rule the dynamic and static properties of nucleons. In specific reference frames, EMFFs represent the Fourier transforms of spatial charge and magnetic momentum densities. 

Recently, the BESIII experiment~\cite{BEPC2site} measured the time-like nucleon form factors (FFs) at center-of-mass energies between 2.0 GeV and 3.5 GeV~\cite{PhysRevD.99.092002, 2021, PhysRevLett.124.042001,YANG20232729,2021136328}. The obtained data present an oscillating behavior~\cite{Bianconi:2015owa,Bianconi:2015vva,Bianconi:2016bss,Bianconi:2018unb,Tomasi-Gustafsson:2020vae,Tomasi-Gustafsson:2021omz}, which manifests itself as a periodic, exponentially dumped component over the typical dipolar carrier, usually identified as the dominant contribution to FFs.  The nature of such an oscillating component is still unknown. The possible explanations can be distinguished into two classes of phenomena: the ones relying on the final state interaction between the baryon and the antibaryon and the others due, instead, to dynamical processes intrinsic to the baryon structure and hence encoded by the EMFFs of nucleons.  

In order to investigate this eventuality we propose a parametrization for the EMFFs defined by considering the nucleons as triplets of collinear quarks lying at light-like distances in the light-front framework~\cite{Brodsky_1998}.	 

The matrix element of the ``+'' component of the hadronic current $J^\mu$, which depends directly on the EMFFs, evaluated between the baryon and antibaryon particle states, can then be expanded using the Lorentz invariance of the three quark Fock state's matrix elements. 

The resulting form depends on a set of functions of the four momentum squared fractions, called light cone distribution amplitudes (LCDAs), and a deep knowledge of their expression can provide further information about the form factors shape. Using the $\mathcal{L}_\text{QCD}$ conformal symmetry~\cite{Braun_1999}, the LCDAs are expanded on a polynomial basis, the most common choice being represented by the orthonormal Appell polynomials, defined on the triangle $T\left(x_1,x_3\right)=\{\left(x_1,x_3\right)\in \mathbb{R}:x_1>0,x_3>0,x_1+x_3<1\}$, where $x_i=k_i^+/P^+$ is the $i^{\rm th}$ quark's light front momentum fractions along is the $\left(^+\right)$ direction, with $i=1,2,3$ and so the following relation holds: $\sum_{i=1}^3x_i=1$. The only unknown quantities now are the expansion coefficients, which have to be determined considering the phenomenology of the reaction. The non-perturbative coefficients admit an evolution equation in the conformal symmetry framework, and their values can be determined theoretically by QCD sum rules. On the other hand, we are considering a center of mass energy of the system between 2.0 GeV and 3.5 GeV, so we are not allowed to use perturbative methods. What we propose then is to perform a truncated Laurent expansion of the non-perturbative coefficients over the negative powers of the four momentum squared, subsequently performing a fit over the recent BESIII experimental data to determine these coefficients. The final goal of this description is to find whether the oscillations of the EMFFs can be described by the model functions.
\section{The microscopic model\label{sec:micro}}
One of the most effective ways to describe subnuclear processes is to work on a light-front
 framework, expanding the involved particle states in a free particle state basis, commonly known as Fock states. For a baryon we have 
\begin{eqnarray}
 \left|{\rm baryon}\right\rangle=\left|0\right\rangle+\left|qqq\right\rangle+\left|qqqg\right\rangle+\left|qqqq\bar q\right\rangle+\dots\,.
\end{eqnarray}
The matrix element between the vacuum and the particle states of the three-quark state can be expanded in a Lorentz series~\cite{CHERNYAK198452}, which for the proton has the form
\textcolor{black}{
\begin{eqnarray}
\label{LCDAs}
&& \left\langle0\right|\varepsilon^{ijk}u^i_\alpha\left(a_1z\right)u^j_\beta\left(a_2z\right)d^k_\gamma\left(a_3z\right)\left|P\right\rangle
\nonumber\\
&=& \frac{1}{4}\Big[ \mathcal{S}_1MC_{\alpha\beta}\left(\gamma_5N^+\right)_\gamma+\mathcal{V}_1\left(\slashed{p}C\right)_{\alpha\beta}\left(\gamma_5N^+\right)_\gamma\Big.
\nonumber\\
&& +\mathcal{P}_1M\left(\gamma_5C\right)_{\alpha\beta}N_\gamma^+
+\mathcal{A}_1\left(\slashed{p}\gamma_5C\right)_{\alpha\beta}N_\gamma^+\nonumber\\
&&\Big. +\mathcal{T}_1\left(i\sigma_{\bot p}C\right)_{\alpha\beta}\left(\gamma^\bot \gamma_5N^+\right)_\gamma+\ldots\Big]\,.
\end{eqnarray}
In the previous equation, $z$ is a light-cone vector, i.e., $z^2=0$, $q_\alpha^i\left(a_1z\right)$ is a quark operator, where $i$ is a colour index and $\alpha$ is a Dirac index. Moreover, $C$ is the charge conjugation matrix, $\slashed{p}=p_\mu\gamma^\mu$, $N^+$ is the plus component of the nucleon spinor, $\sigma^{\mu\nu}=\frac{i}{2}\left[\gamma^\mu,\gamma^\nu\right]$ and $\left|P\right\rangle$ is the nucleon state. The functions $\mathcal{S}_1$, $\mathcal{V}_1$, $\mathcal{P}_1$, $\mathcal{A}_1$, and $\mathcal{T}_1$ are called light-cone distribution amplitudes,} which are functions of the scalar product $P\cdot z$. The dots in Eq.~\eqref{LCDAs} indicate that the expansion has been written explicitly only for twist-3 LCDAs, while the complete expansion includes 24 LCDAs.
 
 Considering now the Fourier transform of the three quark matrix element defined in Eq.~\eqref{LCDAs}, we can find some conditions for the LCDAs, imposing that the isospin  of the nucleon is 1/2. For example, for twist-3 LCDAs, the following equation holds
 \begin{eqnarray}
 2T_1\left(x_1,x_2,x_3\right)&=&
 \left[V_1-A_1\right]\left(x_1,x_3,x_2\right)
 \no+\left[V_1-A_1\right]\left(x_2,x_3,x_1\right)\,,
 \nonumber
 \end{eqnarray}
 which allows to restrict the study of twist-3 LCDAs to a single function, which is chosen to be
 \begin{equation}
 \varphi_N\left(\mathbf{x}\right)=V_1\left(\mathbf{x}\right)-A_1\left(\mathbf{x}\right)\,,
 \nonumber
 \end{equation}
 with the 3-vector $\mathbf{x}=(x_1,x_2,x_3)$. Taking advantage from the conformal symmetry of the Lagrangian density $\mathcal{L}_\text{QCD}$, the twist-3 LCDA $\varphi_N$ can be expanded over the orthonormalized Appell polynomials set $\left\{P_n\left(\mathbf{x}\right)\right\}_n$ as follows
 \begin{equation}
 \varphi_N\left(\mathbf{x},Q^2\right)=120x_1x_2x_3\sum_n B_n\left(Q^2\right)P_n\left(\mathbf{x}\right)\,.
 \nonumber
 \end{equation}
 The set of non-perturbative coefficients $\left\{B_n\right\}_n$ is unknown and contains all the information about the form factor for the leading twist.
The $Q^2$-dependence of the twist-3 LCDA $\varphi_N$ is entirely due to the coefficients of the set $\left\{B_n\right\}_n$. The first coefficient $B_0$ is constant and well-known, being linked to the normalization of $\varphi_N$, i.e.,
 \begin{equation}
 B_0=\int_{0}^{1}dx_1\int_{0}^{1}dx_2\int_{0}^{1}dx_3\varphi_N\left(\mathbf{x},Q^2\right)=1\,.
 \nonumber
 \end{equation}
As already stated, we aim to study the non-perturbative aspects of the LCDAs by considering $Q^2$-dependent coefficients for the Lorentz expansion of $\varphi_N\left(\mathbf{x},Q^2\right)$. Moreover, to make a smooth continuation with the $1/Q^2$-power-law asymptotic behavior of EMFFs predicted by perturbative QCD, which is reached at large values of $|Q^2|$ in space and time-like regions, we consider for tor the Lorentz coefficients an expansion over the negative powers of the four-momentum squared,
\begin{equation}
\label{eq:seriesexpansion}
B_n\left(Q^2\right)=\sum_{k=0}^{M_n}b_k^{\left(n\right)}Q^{-2k},
\end{equation}
where $\left\{\{b_k^{(n)}\}_{k=0}^{M_n}\right\}_{n}$ is the set of coefficients and  $M_n$ is the maximum power of $Q^{-2}$ in the expansion of the $n^{\rm th}$ coefficient $B_n(Q^2)$.
\section{Leading order contributing diagrams}
Taking the leading order into account, the minimum number of contributing diagrams to be considered is fourteen~\cite{CHERNYAK198452}. The corresponding Feynman diagrams and hard scattering kernels (see below) are reported in Table~\ref{Table:CZdiagrams}. The Sachs form factors which parametrize the electromagnetic 4-current of the baryon vertex $B\gamma B$ are related to the Pauli and Dirac ones by the relations
\begin{eqnarray}
 G_E\left(Q^2\right)&=&F_1\left(Q^2\right)-\tau F_2\left(Q^2\right)\,,\nonumber\\
 G_M\left(Q^2\right)&=&F_1\left(Q^2\right) + F_2\left(Q^2\right)\,,\nonumber
\end{eqnarray}
where $\tau=Q^2/\left(4M_B^2\right)$, being $M_B$ the baryon mass. For theoretical and experimental reasons, we analyze the so-called effective form factor, defined in terms of the moduli of Sachs form factors as
\begin{eqnarray}
	G_\text{eff}=\sqrt{\frac{\left|G_E\right|^2+2\tau\left|G_M\right|^2}{1+2\tau}}.
\nonumber\end{eqnarray}
From the theoretical point of view, considering only twist-3 LCDAs, we are assuming the coincidence of the Sachs form factors, this means that their common moduli does equal the effective form factor. In fact, from the previous definition, it follows that  
\begin{equation}
\label{eq:FFAssumption}
G_{\text{eff}}\left(Q^2\right)
=\left|G_E\left(Q^2\right)\right|=\left|G_M\left(Q^2\right)\right|\,.
\end{equation}
Following the works of Brodsky and Lepage~\cite{PhysRevD.22.2157}, the light front EMFF can be written as the convolution of three probability amplitudes, namely: that of having the three valence quarks bounded in the incoming baryon, that of having the same three quarks bounded in the out-coming baryon, and the probability amplitude of finding a certain strong interaction, known as ``hard scattering kernel'' $K_H$. At the leading order, see Table~\ref{Table:CZdiagrams}, each of the fourteen contributing diagrams corresponds to a hard scattering kernel of the set $\left\{K_{i}\left(\mathbf{x},\mathbf{y}\right)\right\}_{i=1}^{14}$, here $\varphi_N\left(\mathbf{x}\right)=V_1\left(\mathbf{x}\right)-A_1\left(\mathbf{x}\right)$, and $T\left(\mathbf{x}\right)=T_1\left(\mathbf{x}\right)$ are the light cone distribution amplitudes involved in the calculation. The hard scattering kernels have been obtained by evaluating the contributing diagrams as shown in the Appendix~\ref{appendix:b}.
\begin{table}[h]
\caption{Feynman diagrams and hard scattering kernels~\cite{CHERNYAK198452}}.\label{Table:CZdiagrams}
	\begin{tabular}{c c c }
	\hline
		Index $i$ & Diagram & $K_{i}(\mathbf{x},\mathbf{y})$ \\
		\hline
		1 & \begin{minipage}{20mm}\includegraphics[width = 20 mm]{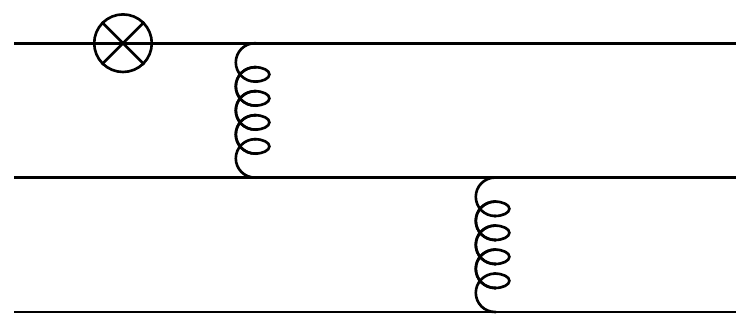}\end{minipage} & $\displaystyle \frac{\varphi_N(\mathbf{x})\varphi_N(\mathbf{y})+4T(\mathbf{x})T(\mathbf{y})}{(1-x_1)^2 x_3(1-y_1)^2 y_3}$\\
		2 & \begin{minipage}{20mm}\includegraphics[width = 20 mm]{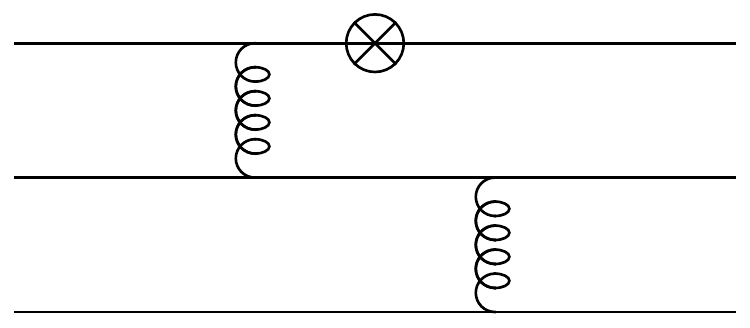}\end{minipage} & 0 \\
		3 & \begin{minipage}{20mm}\includegraphics[width = 20 mm]{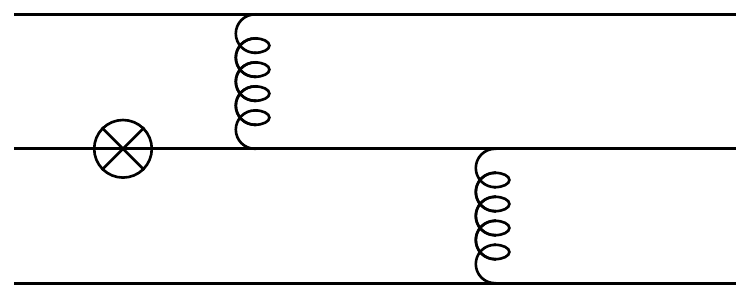}\end{minipage} & $\displaystyle \frac{-4T(\mathbf{x})T(\mathbf{y})}{x_1x_3(1-x_2)y_1y_3(1-y_1)}$\\
		4 & \begin{minipage}{20mm}\includegraphics[width = 20 mm]{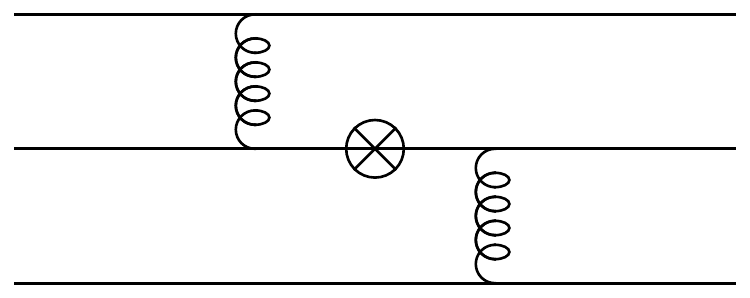}\end{minipage} & $\displaystyle \frac{\varphi_N(\mathbf{x})\varphi_N(\mathbf{y})}{x_1x_3(1-x_3) y_1y_3(1-y_1)}$\\
		5 & \begin{minipage}{20mm}\includegraphics[width = 20 mm]{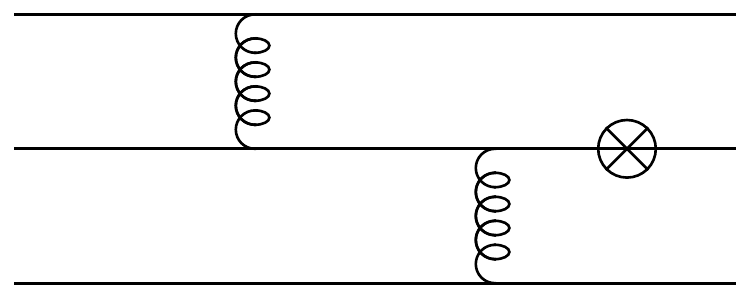}\end{minipage} & $\displaystyle \frac{-\varphi_N(\mathbf{x})\varphi_N(\mathbf{y})}{x_2x_3(1-x_3) y_2y_3(1-y_1)}$\\
		6 & \begin{minipage}{20mm}\includegraphics[width = 20 mm]{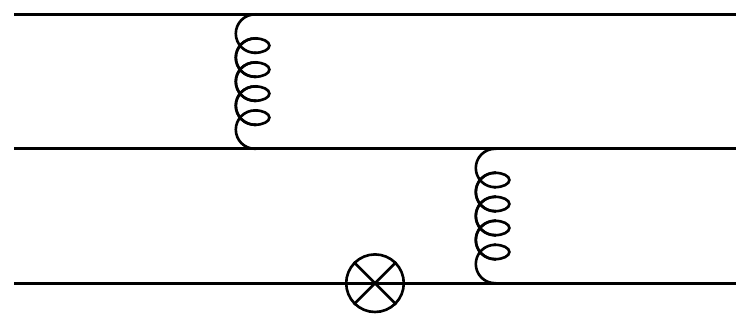}\end{minipage} & 0 \\
		7 & \begin{minipage}{20mm}\includegraphics[width = 20 mm]{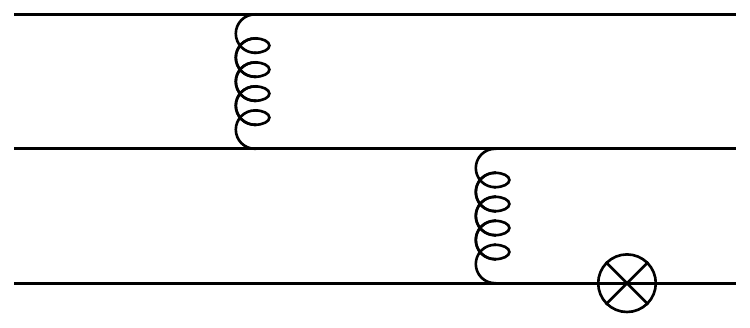}\end{minipage} & $\displaystyle \left(\frac{1}{x_1y_1}+\frac{1}{x_2y_2}\right)\frac{\varphi_N(\mathbf{x})\varphi_N(\mathbf{y})}{(1-x_3)^2(1-y_3)^2}$ \\
		8 & \begin{minipage}{20mm}\includegraphics[width = 20 mm]{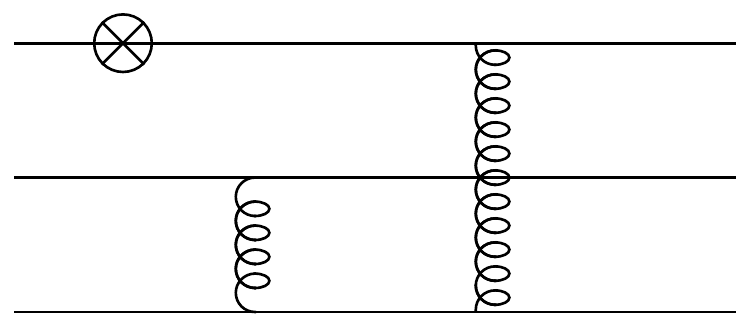}\end{minipage} & 0 \\
		9 & \begin{minipage}{20mm}\includegraphics[width = 20 mm]{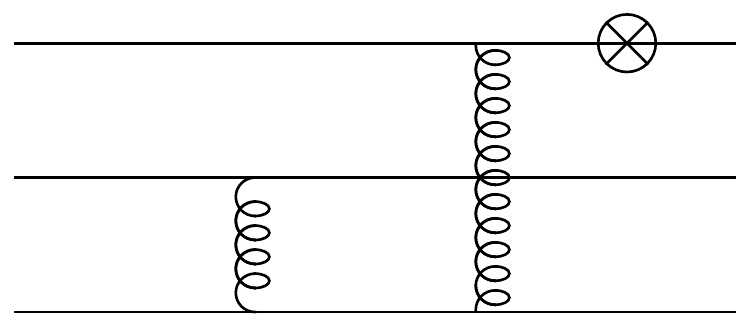}\end{minipage} & $\displaystyle \frac{\varphi_N(\mathbf{x})\varphi_N(\mathbf{y})+4T(\mathbf{x})T(\mathbf{y})}{(1-x_1)^2 x_2(1-y_1)^2 y_2}$\\
		10 & \begin{minipage}{20mm}\includegraphics[width = 20 mm]{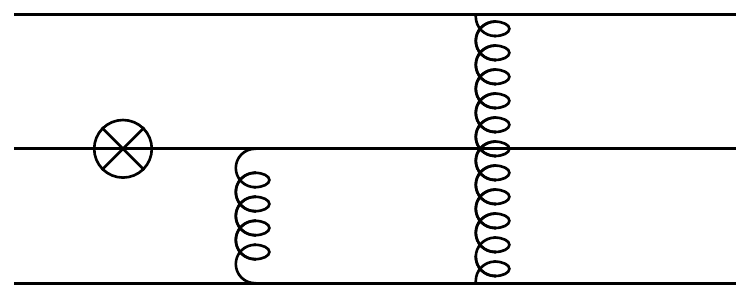}\end{minipage} & $\displaystyle \frac{\varphi_N(\mathbf{x})\varphi_N(\mathbf{y})+4T(\mathbf{x})T(\mathbf{y})}{(1-x_1)^2 x_2(1-y_1)^2 y_2}$\\
		11 & \begin{minipage}{20mm}\includegraphics[width = 20 mm]{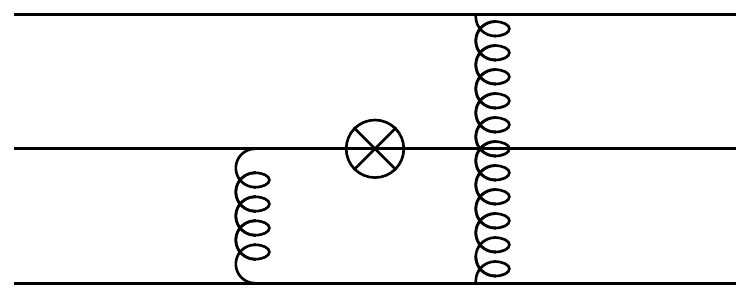}\end{minipage} & 0 \\
		12 & \begin{minipage}{20mm}\includegraphics[width = 20 mm]{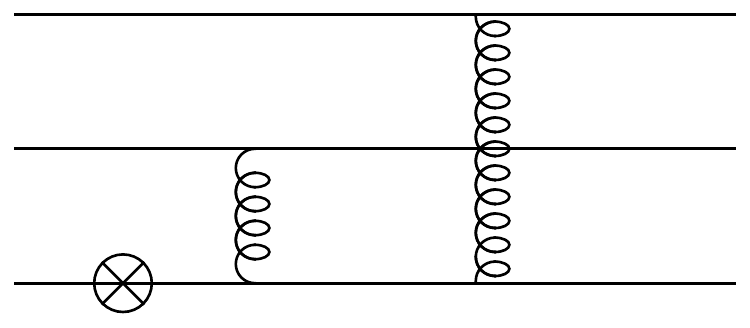}\end{minipage} & $\displaystyle \frac{-\varphi_N(\mathbf{x})\varphi_N(\mathbf{y})}{x_1x_2(1-x_3) y_1y_2(1-y_1)}$\\
		13 & \begin{minipage}{20mm}\includegraphics[width = 20 mm]{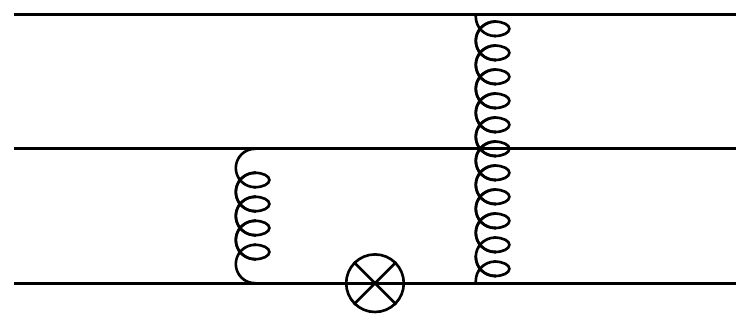}\end{minipage} & $\displaystyle \frac{4T(\mathbf{x})T(\mathbf{y})}{x_1x_2(1-x_1) y_1y_2(1-y_2)}$\\
		14 & \begin{minipage}{20mm}\includegraphics[width = 20 mm]{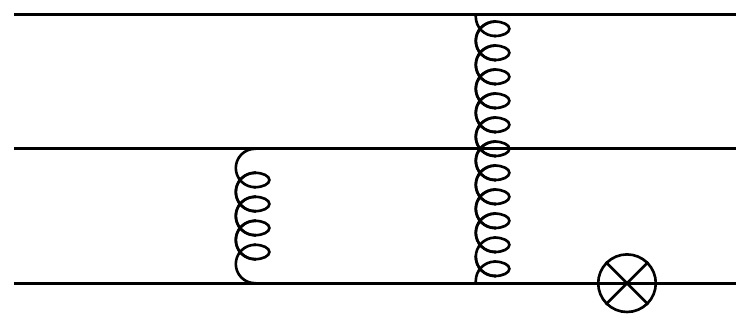}\end{minipage} & $\displaystyle \frac{-\varphi_N(\mathbf{x})\varphi_N(\mathbf{y})}{x_1x_2(1-x_1) y_1y_2(1-y_3)}$\\
	\end{tabular}
\end{table}\\
To evaluate the effective form factor $G_{\rm eff}$, we use the asymptotic formula of Chernyak-Zhitnitsky~\cite{CHERNYAK198452}, which, including non-perturbative effects at low $|Q^2|$, by considering the $Q^2$-dependent Lorentz coefficient of Eq.~\eqref{eq:seriesexpansion}, can be interpreted as an identity, so we have
\begin{eqnarray}
\label{eq:CZ}
&\displaystyle 
G_{\rm eff}\left(q^2\right)\blu{=} \frac{\left(4\pi\bar\alpha_s\right)^2}{54(q^2)^2}\left|f_N\right|^2
&\nonumber\\
&\displaystyle \times\!\!\!
\int\!\!\left[dx\right]\!\!
\int\!\!\left[dy\right]\!\left(\!\!2\sum_{i=1}^{7}e_iK_i\left(\mathbf{x},\mathbf{y}\right)+\!\sum_{i=8}^{14}e_iK_i\left(\mathbf{x},\mathbf{y}\right)\right)\,,&
\end{eqnarray}
 where: $f_N$ is the value of nucleon wave function at the origin, $\left[dx\right]=\delta\left(1-\sum_{i=1}^{3}x_i\right)dx_1dx_2dx_3$ and $\bar \alpha_s$ is the modified strong coupling constant. 
 The value of $\bar \alpha_s^2$ is given by the product of the coupling constants for the two subprocesses~\cite{CHERNYAK198452}, namely the two gluon exchanges which appear in the tree-level diagrams shown in Table~\ref{Table:CZdiagrams}. The average virtuality ${\bar q}_1^2$ of the lightest gluon is ${\bar x}_3{\bar y}_3q^2$, while the second gluon has an averaged virtuality ${\bar q}_2^2=\overline{(1-x_1)}\,\,\overline{(1-y_1)}q^2$. The typical values of a realistic nucleon wave function for the momentum fractions are ${\bar x}_1\simeq 2/3$ and ${\bar x}_2\simeq {\bar x}_3\simeq 1/6$, therefore $\bar \alpha_s^2\left(q^2\right)=\alpha_s\left(\bar{q}^2_1\right)\alpha_s\left(\bar{q}_2^2\right)=\alpha_s\left(q^2/36\right)\alpha_s\left(q^2/9\right)$.
 The integrals in Eq.~\eqref{eq:CZ} are weakly convergent, it is possible to solve them analytically, the detailed computation is reported in Appendix~\ref{appendix:a}. The results have already been obtained in Ref.~\cite{Braun_1999}, and are reported in Appendix~\ref{appendix:c}. 
\\\indent
As it is well known, QCD gives prescriptions for asymptotics, even though the asymptotic threshold, i.e., the $q^2$ value from which such prescriptions apply can not be defined. Moreover, the question of their validity in the time-like region experimentally accessible, which starts from the production threshold $q^2=(2M_B)^2$, is especially interesting and controversial. The two-nucleon threshold region, $(2M_N)^2\sim 3.5$ GeV$^2$, on one side corresponds to the small distances characteristic of QCD below $0.1$ fm, on the other hand, it concerns systems of particle and anti-particle having energies so small that can be treated in terms of classical quantum mechanics.
\\\indent
In our opinion, both approaches (perturbative and non-perturbative) are relevant for specific aspects~\cite{TOMASIGUSTAFSSON2001291,Tomasi_Gustafsson_2005}. Indeed, the oscillations found in Ref.~\cite{Bianconi:2015owa} are typically of a size of 10\% with respect to the perturbative QCD background, so that the three valence quark configuration strongly contributes to the nucleon structure.
\\\indent
Final state interaction of baryon and anti-baryon, namely re-scattering processes in $e^+e^-\to B\bar{B}$, discussed, for instance, in Ref.~\cite{YANG20232729}, have been not considered, because this study aims to verify whether a three quark description of the nucleon-antinucleon production process was sufficient to reproduce the experimental data. For that reason, we applied a model which takes only the nucleons' internal structure into account, studying the discrepancy from the monotone background, which can be well described by a time-like dipole interaction. The authors of Ref.~\cite{YANG20232729} consider final state interaction specifically, which, if included, would go against our purpose of describing the oscillations as a purely nucleons' internal structure effect.
\\\indent
As already stated, non-perturbative-QCD effects are accounted for by considering $Q^2$-dependent coefficients $\left\{B_n(Q^2)\right\}_n$, defined in Eq.~\eqref{eq:seriesexpansion} as truncated expansions in powers of $Q^{-2}$. The first coefficient $B_0$, related to the zero order of the LCDA $\varphi_N$, is constant and fixed to 1. As for the other parameters, we are for now limiting our discussion to the LCDA second order momenta, so we are only interested in the first six coefficients, namely those of the set $\{B_n\}_{n=0}^{5}$.

For the truncated expansions, we propose $M_0=0$, $M_1=M_2=1$, $M_3=M_4=M_5=2$, so that: $B_0=1$ and
\begin{eqnarray}
B_n(Q^2)&=& b_0^{(n)}+\frac{b_1^{(n)}}{Q^2}\,,\ \ \ n=1,2\,,
\nonumber\\ B_m(Q^2) &=& b_0^{(m)}+\frac{b_1^{(m)}}{Q^2}+\frac{b_2^{(m)}}{(Q^2)^2}\,,\ \ \ m=3,4,5\,.
\label{eq:parameters}
\end{eqnarray}
We obtain a closed expression of the effective form factor depending only on the non-perturbative parameters for the proton and the neutron. As already stated in Ref.~\cite{Braun_1999}, since the nucleons are related by the isospin symmetry, and the colour structure of the neutron three quark matrix element is the same as the proton one, their parameters are the same and the effective form factors defined by Eq.~\eqref{eq:CZ} differ only by the values of the electric charges of the quarks which interact with the virtual photon. Due to this property, we performed a simultaneous fit to the recent BESIII data on proton and neutron cross sections to determine the sets of coefficients $\big\{\big\{b_k^{(n)}\big\}_{k=0}^{M_n}\big\}_{n=0}^5$. The fit is performed using the ROOT Data Analysis Framework~\cite{BRUN199781} program.
\section{Results and discussion}
\begin{figure}[h]
\begin{center}
	\includegraphics[width=\columnwidth]{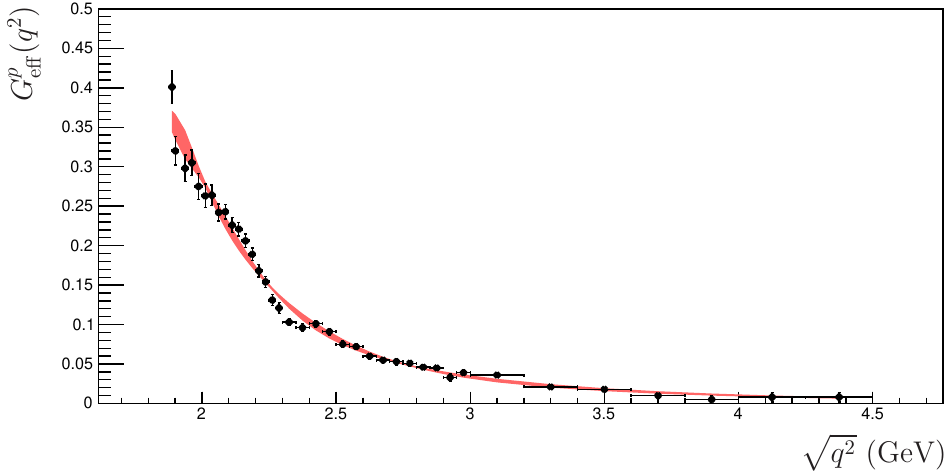}
		\\
	\includegraphics[width=\columnwidth]{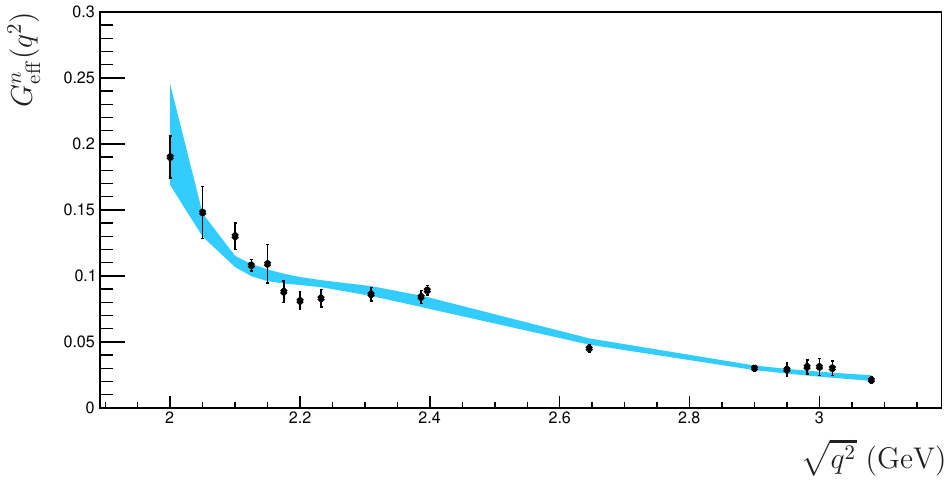}
	\caption{The bands represent the fit results for the proton (upper panel) and neutron (lower panel) effective form factor. The data are from the BESIII experiment~\cite{PhysRevD.99.092002, 2021, PhysRevLett.124.042001}.\label{fig:results}}
	\end{center}
\end{figure}

Figure~\ref{fig:results} shows the results for the effective proton (upper panel) and neutron (lower panel) form factors in comparison with the experimental values measured by the  BESIII experiment~\cite{PhysRevD.99.092002, 2021, PhysRevLett.124.042001}. 
We use only the BESIII data because they are the most suitable to test our model, having small errors, starting just at the production thresholds and covering wide $q^2$ regions.
The fit functions depend on 13 free parameters, which are the coefficients of the expressions of Eq.~\eqref{eq:parameters} and are reported with the corresponding errors in Table~\ref{Table:Params}.
\begin{table}[h]
\caption{\textcolor{black}{Best values of fit parameters with errors.\label{Table:Params}}}
\begin{tabular}{c | c | r}
\hline
\textbf{Parameter} & \textbf{Coefficient} & \textbf{Value} \\
\hline
\multirow{2}{*}{$B_1$} & $b_0^{\left(1\right)}$ & $-31\pm 1$\\
 & $b_1^{\left(1\right)}$ & $\left(-144\pm11\right)$ GeV$^2$\\
\hline
\multirow{2}{*}{$B_2$} & $b_0^{\left(2\right)}$ & $-15.3\pm0.1$\\
 & $b_1^{\left(2\right)}$ & $\left(36\pm1\right)$ GeV$^2$\\
\hline
\multirow{3}{*}{$B_3$} & $b_0^{\left(3\right)}$ & $7.2\pm0.3$\\
 & $b_1^{\left(3\right)}$ & $\left(-80\pm3\right)$ GeV$^2$\\
 & $b_2^{\left(3\right)}$ & $\left(251\pm7\right)$ GeV$^4$\\
\hline
\multirow{3}{*}{$B_4$} & $b_0^{\left(4\right)}$ & $56\pm2$\\
 & $b_1^{\left(4\right)}$ & $\left(-2.95\pm0.03\right)\cdot 10^3$ GeV$^2$\\
 & $b_2^{\left(4\right)}$ & $\left(6.7\pm0.1\right)\cdot 10^3$ GeV$^4$\\
\hline
\multirow{3}{*}{$B_5$} & $b_0^{\left(5\right)}$ & $0.6\pm0.7$\\
 & $b_1^{\left(5\right)}$ & $\left(80\pm6\right)$ GeV$^2$\\
 & $b_2^{\left(5\right)}$ & $\left(-135\pm21\right)$ GeV$^4$\\
\end{tabular}
\end{table}\\
The normalized minimum $\chi^2$ is
\begin{eqnarray}
	\frac{\chi^2}{n_{\rm DoF}}=\frac{76.93}{56-13}\simeq 1.79\,.
\nonumber\end{eqnarray}
It has been obtained by using 56 data points, 48 of the proton cross section and 18 of the neutron one. The error bands have been determined by considering both the errors of data, and the theoretical systematic error of the model, which has ben estimated by using expressions for the coefficients of the set $\left\{B_n(Q^2)\right\}$ with the additional power $\left(Q^{-2}\right)^{M_n+1}$, with $n=1,2,3,4,5$, see Eq.~\eqref{eq:parameters}. 
\\
\begin{figure}[h]
\begin{center}
	\includegraphics[width=\columnwidth]{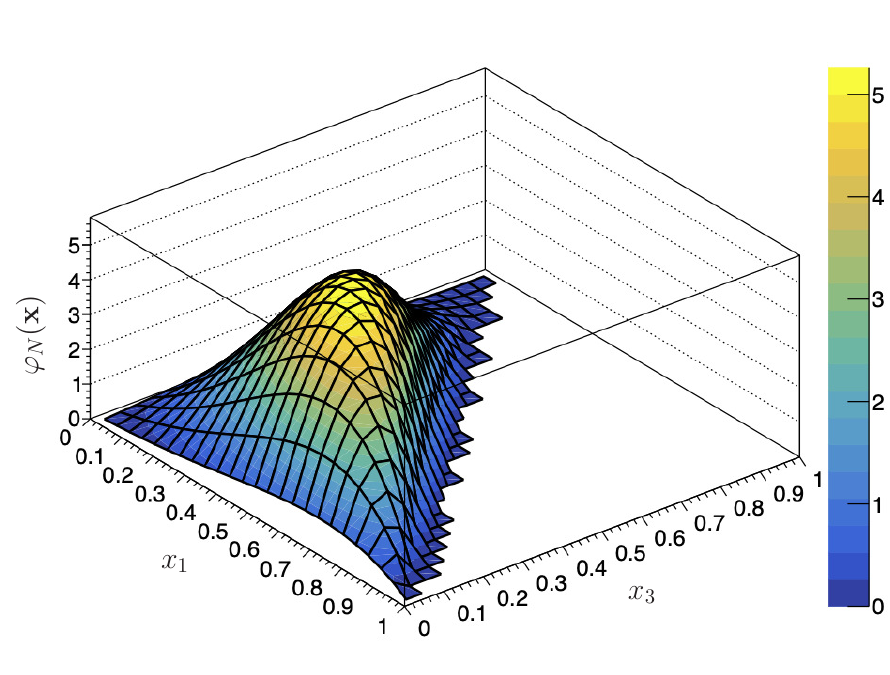}
	\end{center}
	\caption{The nucleon distribution function $\varphi_N(\mathbf{x})$ obtained at $\sqrt{-Q^2}=2.5$ GeV as a function of $x_1$ and $x_3$.\label{fig:phi-n}}
\end{figure}\\
Figure~\ref{fig:phi-n} shows the twist-3 nucleon distribution amplitude $\varphi_N(\mathbf{x},Q^2)$ evaluated at $\sqrt{-Q^2}=2.5$ GeV. 
This value has been chosen to make a comparison with the Chernyak-Zhitnitsky model~\cite{CHERNYAK198452}. In fact, in that case the $Q^2$-dependence of the LCDA was only due the running of the strong coupling constant, while the coefficients of the set $\left\{B_n\right\}_n$  are kept constant to their values at $\sqrt{-Q^2}\sim 2.5$ GeV.
The maximum value of $\varphi_N(\mathbf{x},Q^2)$ is reached at the light-cone momentum fractions
\begin{equation}
x_1\simeq 0.49\,, \qquad x_2\simeq0.24\,, \qquad x_3\simeq0.27\,,\nonumber
\end{equation}
which agree with the assumption made in the Chernyak-Zhitnitsky formula~\cite{CHERNYAK198452}, for which the first valence quark has a momentum fraction about 50\% larger than the other two which equally divide the remaining momentum fraction.

Summarizing, a coherent model has been developed to reproduce the data on proton and neutron EMFFs, recently obtained by the BESIII collaboration. The model is based on a parametrization of the light-cone distribution amplitudes, and obeys conformal symmetry of the QCD Lagrangian.

The light-front quantization allows us to express the three-quark operator, occurring in the expression of the hard scattering kernel, as the Fourier transform of the light-cone distribution amplitudes, which, in turn, describe the behaviour of the three valence quarks constituting the baryon in a light-cone system. Always under the aegis of the QCD-Lagrangian conformal symmetry we expanded the leading-twist baryon distribution amplitudes over a set of Appell polynomials, which diagonalize the one gluon exchange kernel. The non-perturbative nature of the baryon distribution amplitudes is implemented by considering a $Q^2$-dependence of the expansion coefficients of the set $\left\{B_n\right(Q^2)\}_{n}$. We have restricted our calculations to the first six Appell polynomials.

These distribution amplitudes have been used to calculate the near threshold behaviour of the nucleon effective form factors, extending the valence of an asymptotic formula to the low-momentum transfer region. Such a near-threshold extension has been obtained by parametrizing the expansion coefficients as polynomials of zero, first and second degree of $Q^{-2}$. 

Moreover, since the nucleon EMFFs are linked by the isospin symmetry, the same set $\left\{B_n(Q^2)\right\}_{n}$ can be used and the free coefficients of their power series can be determined by means a simultaneous fit to the proton and neutron data. The error bands of the obtained effective nucleon form factors have been obtained by considering the experimental uncertainties of the data, used to determine the free parameters of the model and the theoretical systematic uncertainties due to the particular parametrization used.

This study aims to identify the origin or at least the dominant cause of the oscillatory behaviour of the effective nucleon form factors. In particular, we would like to distinguish between two hypotheses about the oscillation phenomenon: the intrinsic dynamical origin and the final state interaction. In the first case, the oscillations appear at the form factor level, while in the second case, they are due to re-scattering reactions that happen when the nucleons are already formed. A necessary condition in favour of the intrinsic origin is that there exists a microscopic model of nucleons that can give oscillatory form factors. 

Even though for the neutron effective form factor the model reproduces quite well the oscillatory behaviour, it seems to fail in the case of the proton. Indeed, the obtained behaviour of the effective proton form factor, the orange band shown in the upper panel of Fig.~\ref{fig:results}, is compatible with the so-called regular background of Refs.~\cite{Bianconi:2015owa,Bianconi:2015vva,Tomasi-Gustafsson:2020vae}. It can be interpreted as the contribution due to the short distance quark-level dynamics~\cite{Matveev:1973ra,Brodsky:1973kr}, i. e., the $p\bar p$ final state is produced by the creation of quark-antiquark pairs within a small volume, with a linear dimension much smaller than the standard hadron size of about 1 fm. 

Nevertheless, the model has the added value of proving that a unique parametrization in all the kinematical ranges where data are present is effective both for proton and neutron form factors. This is in contrast to previous works, where a common fit could only be achieved either in a restricted kinematical region, concluding in a change of the phase~\cite{2021}, or at the price of three different models applicable in different kinematical regions~\cite{YANG20232729}.

\newpage
\appendix
\section{Contributing diagrams computation}
\label{appendix:a}
Using the expressions~\cite{CHERNYAK198452}
\begin{eqnarray}
\varphi_N(\mathbf{x})
&=&
120 x_1x_2x_3\left( a x_1^2+bx_2^2+cx_3^2+dx_3+e\right)	\,,
\n
T(\mathbf{x})
&=&
120 x_1x_2x_3\left[ \frac{a+c}{2}\left( x_1^2+x_2^2\right)+bp_3^2+\right.
\no\left.\frac{d}{2}(1-x_3)+e\right]	\,,
\end{eqnarray}
where the values of the coefficients $a$, $b$, $c$, $d$, $e$ and $f$ are given in Ref.~\cite{CHERNYAK198452}, the analytic solutions of the integrals of Eq.~\eqref{eq:CZ}, namely the ten non-vanishing expressions
\begin{eqnarray}
	{\cal K}_i=\int[dx]\int[dy] K_i(\mathbf{x},\mathbf{y})\,,
\end{eqnarray}
where the functions $K_i(\mathbf{x},\mathbf{y})$ are given in Table~\ref{Table:CZdiagrams}, with $i\in\{1,3,4,5,7,9,10,12,13,14\}$, are:
\begin{eqnarray}
{\cal K}_1
&=&\left(I_1^{\left(1\right)}\right)^2+4\left(I_1^{\left(2\right)}\right)^2\,,
\nonumber\\
I_1^{(1)}&=&
 \frac{5}{12}(36 a + 6b + 2c + 12 d + 72e)\,,
\nonumber\\
 I_1^{\left(2\right)}& =& 
 \frac{5}{12} (21 a+20 b+3 c+32 d+72 e)\,;
\\
%
{\cal K}_3 &=& -4I_3^{(1)}I_3^{(2)}\,,	
\nonumber\\
I^{(1)}_3 &=&
\frac{5}{3} (10 a+2 b+10 c+15 d+36 e)\,,
\nonumber\\
I^{(2)}_3&=&
\frac{5}{12} (15 a+6 b+15 c+28 d+72 e)\,;
\\
%
	{\cal K}_4 &=& I_4^{(1)}I_4^{(2)}\,,
	\n
	I_4^{(1)}&=&
	 \frac{5}{2} (a+3 b+2 (c+2 d+6 e))\,,
	 \n
	 I^{(2)}_4
	 &=&
	 \frac{5}{6} (6 a+9 b+3 c+8 d+36 e)\,;
\\
%
{\cal K}_5 &=& -I_5^{(1)}I_5^{(2)}\,,
\n
I_5^{(1)}
&=&
 \frac{5}{2} (3 a+b+2 (c+2 d+6 e))
 \n
 I^{(2)}_5 
 &=&
\frac{10}{3} (9 a+b+c+3 d+18 e)\,;
\\
%
	{\cal K}_7&=& \left(I^{(1)}_7\right)^2+\left(I^{(2)}_7\right)^2\,,
	\n
	I^{(1)}_7 
	&=&
	\frac{5}{6} (a+3 (b+6 c+8 d+12 e))\,,
	\n
	I^{(2)}_7 
	&=&
\frac{5}{6} (3 a+b+6 (3 c+4 d+6 e))\,;
\\
%
	{\cal K}_9 &=& {\cal K}_{10} = \left(I^{(1)}_9\right)^2+4\left(I^{(2)}_9\right)^2\,,
	\n
	I^{(1)}_9 
	&=&
	\frac{5}{6} (18 a+b+3 c+8 d+36 e)\,,
	\n
	I^{(2)}_9 
	&=&
	 \frac{5}{12} (19 a+6 b+19 c+28 d+72 e)\,;
\end{eqnarray}
\begin{eqnarray}
	{\cal K}_{12} 
	&= & - I_{12}^{(1)}I_{12}^{(2)} \,;
\n
I_{12}^{(1)}
&=&
\frac{10}{3} (a+b+9 c+12 d+18 e)\,,
\n
I_{12}^{(2)}
&=&
\frac{5}{6} (6 a+3 b+9 c+16 d+36 e)\,;
\\
%
{\cal K}_{13}
&=&	4I_{13}^{(1)}I_{13}^{(2)}
\n
I_{13}^{(1)} 
&=& 
\frac{5}{12} (-3 a+90 b-3 c+100 d+360 e)\,,
\n
I_{13}^{(2)} 
&=&
\frac{5}{12} (9 a+18 b+9 c+20 d+72 e)\,;
\\
%
	{\cal K}_{14} &=&
	-I_{14}^{(1)}I_{14}^{(2)}\,,
	\n
	I_{14}^{(1)}
	&=&
	\frac{5}{6} (6 a+3 b+9 c+16 d+36 e)\,,
	\n
	I_{14}^{(2)}
	&=&
	 \frac{10}{3} (a+b+9 c+12 d+18 e)\,.
\end{eqnarray}
%
%
%
%
\section{Distribution amplitudes computation}
\label{appendix:b}
\subsection{Quark distribution amplitudes complex conjugate computation}
\begin{figure}[h]
\begin{center}
	\includegraphics[width= .7\columnwidth]{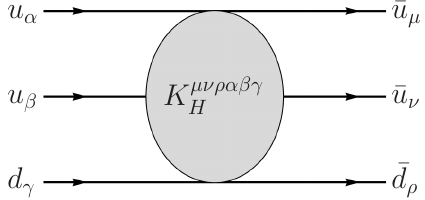}
	\caption{Basic graph for the hard scattering kernel.}
\label{fig:kernel}
\end{center}	
\end{figure}

Omitting the color and current indices, there is a proportionality between the hadronic current matrix element and the quark distribution amplitudes, given by
\begin{equation}
\left\langle P'\right|\!J\!\left|P\right\rangle\propto \left\langle P'\right|\!\bar u_\mu\bar u_\nu \bar d_\rho\!\left|0\right\rangle\left\langle0\right|\!K_H^{\mu\nu\rho\alpha\beta\gamma}\!\left|0\right\rangle\left\langle 0\right|\!u_\alpha u_\beta d_\gamma\!\left|P\right\rangle,
\end{equation}
so that the transition amplitude is proportional to the vacuum expectation value of the hard scattering kernel. The Feynman diagram is depicted in Fig.~\ref{fig:kernel}.
\\
The hard scattering structure $K_H^{\mu\nu\rho\alpha\beta\gamma}$ is given by a product of the three Lorentz structures which give the coupling of the three fermion lines:
\begin{equation}
K_H^{\mu\nu\rho\alpha\beta\gamma}=\Gamma_1^{\alpha\mu}\Gamma_2^{\beta\nu}\Gamma_3^{\gamma\rho}\,.
\end{equation}
First of all, we need to write the complex conjugate of the matrix element representing the distribution amplitude $\varphi_N\left(\mathbf{x}\right)$. We calculate this directly as we can take the complex conjugate of the three parts composing the matrix element expansion:
\begin{eqnarray}
\left\langle0\right|u_\alpha u_\beta d_\gamma\left|P\right\rangle^*
&=&
\left\langle P\right|d^\dagger_\gamma u^\dagger_\beta u^\dagger_\alpha\left|0\right\rangle
\nou
\left\langle P\right|\bar d_i \bar u_j \bar u_k\left|0\right\rangle \gamma^0_{i\gamma}\gamma^0_{j\beta}\gamma^0_{k\alpha}\,.
\end{eqnarray}
For the axial component we have
\begin{eqnarray}
\left\langle P\right|\bar d_i \bar u_j \bar u_k\left|0\right\rangle_A
&=&
A_1^*\left(\slashed{P}\gamma_5C\right)^*_{\alpha\beta}N\left(P\right)^*_\gamma\gamma^0_{\gamma i}\gamma^0_{\beta j}\gamma^0_{\alpha k}
\nou
A_1^*\left(\left(\slashed{P}\gamma_5C\right)^\dagger\right)_{\beta\alpha}\bar N\left(P\right)_i\gamma^0_{\beta j}\gamma^0_{\alpha k}
\nou
A_1^*\left(C\gamma_5\slashed{P}\right)_{jk}\bar N\left(P\right)_i\,,
 \end{eqnarray}
for the vector one
\begin{eqnarray}
\left\langle P\right|\bar d_i \bar u_j \bar u_k\left|0\right\rangle_V
&=&
V_1^*\left(\slashed{P}C\right)^*_{\alpha\beta}\left(\gamma_5N\left(P\right)\right)^*_\gamma\gamma^0_{\gamma i}\!\gamma^0_{\beta j}\gamma^0_{\alpha k}
\nou
-V_1^*\left(\left(\slashed{P}C\right)^\dagger\right)_{\beta\alpha}\left(\bar N\left(P\right)\gamma_5\right)_i\gamma^0_{\beta j}\gamma^0_{\alpha k}
\nou
-V_1^*\left(C\slashed{P}\right)_{jk}\left(\bar N\left(P\right)\gamma_5\right)_i\,,
\end{eqnarray}
and finally for the tensorial part
\begin{eqnarray}
\left\langle P\right|\bar d_i \bar u_j \bar u_k\left|0\right\rangle_T
&=&
T_1^*\left(\tensor{\sigma}^{\mu\nu}P_\nu C\right)^*_{\alpha\beta}\left(\gamma_\mu\gamma_5N\left(P\right)\right)^*_\gamma
\no
\gamma^0_{\gamma i}\gamma^0_{\beta j}\gamma^0_{\alpha k}
\nou
-T_1^*\left(\left(\tensor{\sigma}^{\mu\nu}P_\nu C\right)^\dagger\right)_{\beta\alpha}\left(\bar N\left(P\right)\gamma_5\gamma_\mu\right)_i
\no
\gamma^0_{\beta j}\gamma^0_{\alpha k}
\nou
-T_1^*\left(CP_\nu\tensor{\sigma}^{\mu\nu}\right)_{jk}\left(\bar N\left(P\right)\gamma_5\gamma_\mu\right)_i\,.
\no\end{eqnarray}
\subsection{Distribution amplitudes convolutions}
Here we compute the convolution of the distribution amplitudes used for the evaluation of the contributing diagrams. In each of the following calculations we use the anti-commutative properties of the gamma matrices, in particular the fact that the $\gamma_5$ matrix anti commutes with every component of the matrix four-vector $\gamma^\mu$, while for the charge conjugation matrix $C$ the following rule holds:
\begin{equation}
C\gamma^\mu C=-\left(\gamma^\mu\right)^T\,.
\end{equation}
The axial, vector, tensor, and axial-vector components of the hadronic-current matrix element are, respectively,
\begin{eqnarray}
\label{eq:axialcoupling}
\left\langle P'\right|\!J\!\left|P\right\rangle_A
\!\!&=& \!
\left\langle P'\right|\!\bar u_\mu\bar u_\nu \bar d_\rho\!\left|0\right\rangle_A\!\left\langle0\right|\!K_H^{\mu\nu\rho\alpha\beta\gamma}\!\left|0\right\rangle\!
\left\langle 0\right|\!u_\alpha u_\beta d_\gamma\!\left|P\right\rangle_A
\nou
A_1^*\left(C\gamma_5\slashed{P}'\right)_{\mu\nu}\left(\bar N P'\right)_\rho
\Gamma_1^{\mu\alpha}
\Gamma_2^{\nu\beta}
\Gamma_3^{\rho\gamma}
A_1
\no
\left(\slashed{P}\gamma_5C\right)_{\alpha\beta}N\left(P\right)_\gamma
\nou
\left|A_1\right|^2\bar N\left(P'\right)\Gamma_3N\left(P\right)\no
\left[-\Gamma_1^{\mu\alpha}\left(\slashed{P}\gamma_5C\right)_{\alpha\beta}\left(\Gamma_2^T\right)^{\beta\nu}\left(C\gamma_5\slashed{P}'\right)_{\nu\mu}\right]
\nou
-\left|A_1\right|^2\bar N\left(P'\right)\Gamma_3N\left(P\right)\no
\text{Tr}\left[\Gamma_1\slashed{P}\gamma_5C\Gamma_2^TC\gamma_5\slashed{P}'\right]
\nou
\left(-1\right)^{1+2n_2}\left|A_1\right|^2\bar N\left(P'\right)\Gamma_3N\left(P\right)
\no
\text{Tr}\left[\Gamma_1\slashed{P}\overset{\leftrightarrow}{\Gamma}_2\slashed{P}'\right]\,;
\end{eqnarray}
\begin{eqnarray}
\left\langle P'\right|\!J\!\left|P\right\rangle_V
\!\!&=&\!
 \left\langle P'\right|\!\bar u_\mu\bar u_\nu \bar d_\rho\!\left|0\right\rangle_V\!
 \left\langle0\right|\!K_H^{\mu\nu\rho\alpha\beta\gamma}\!\left|0\right\rangle\left\langle 0\right|\!u_\alpha u_\beta d_\gamma\!\left|P\right\rangle_V
 \nou
 -V_1^*\left(C\slashed{P}'\right)_{\mu\nu}\left(\bar N\left(P'\right)\gamma_5\right)_\rho
 \Gamma_1^{\mu\alpha}\Gamma_2^{\nu\beta}\Gamma_3^{\rho\gamma}
 \no
 V_1\left(\slashed{P}C\right)_{\alpha\beta}\left(N\left(P\right)\gamma_5\right)_\gamma
 \nou
 \left(-1\right)^{1+n_3}\left|V_1\right|^2\bar N\left(P'\right)\Gamma_3N\left(P\right)
 \no
 \left[\left(\Gamma_1\right)_{\mu\alpha}\left(\slashed{P}C\right)_{\alpha\beta}\left(\Gamma_2^T\right)_{\beta\nu}\left(C\slashed{P}'\right)_{\nu\mu}\right]
 \nou
 \left(-1\right)^{1+n_3}\left|V_1\right|^2\bar N\left(P'\right)\Gamma_3N\left(P\right)
 \no
 \text{Tr}\left[\Gamma_1\slashed{P}C\Gamma_2^TC\slashed{P}'\right]
 \nou
 \left(-1\right)^{1+n_2+n_3}\left|V_1\right|^2\bar N\left(P'\right)\Gamma_3N\left(P\right)
 \no
 \text{Tr}\left[\Gamma_1\slashed{P}\overset{\leftrightarrow}{\Gamma}_2\slashed{P}'\right]
 \nou
 -\left|V_1\right|^2\bar N\left(P'\right)\Gamma_3N\left(P\right)\text{Tr}\left[\Gamma_1\slashed{P}\overset{\leftrightarrow}{\Gamma}_2\slashed{P}'\right]\,;
\end{eqnarray}
\begin{eqnarray}
\left\langle P'\right|\!J\!\left|P\right\rangle_T
\!\!&=&\!\!
\left\langle P'\right|\!\bar u_\mu\bar u_\nu \bar d_\rho\!\left|0\right\rangle_T\!\left\langle0\right|\!K_H^{\mu\nu\rho\alpha\beta\gamma}\!\left|0\right\rangle\!\left\langle 0\right|\!u_\alpha u_\beta d_\gamma\!\left|P\right\rangle_T
\nou
T_1^*\left(CP'_\lambda \tensor{\sigma}^{\delta\lambda}\right)_{\mu\nu}\left(\bar N\left(P'\right)\gamma_5\gamma_\delta\right)_\rho
\Gamma_1^{\mu\alpha}
\Gamma_2^{\nu\beta}
\Gamma_3^{\rho\gamma}
\no
T_1\left(\tensor{\sigma}^{\eta\kappa}P_\kappa C\right)_{\alpha\beta}\left(\gamma_\eta\gamma_5\left(N\left(P\right)\right)\right)_\gamma
\nou\left|T_1\right|^2P_\lambda'P_\kappa\bar N\left(P'\right)\gamma_\delta\Gamma_3\gamma_\eta N\left(P\right)
\no
\left[\Gamma_1^{\mu\alpha}\left(\tensor{\sigma}^{\eta\kappa}C\right)_{\alpha\beta}\left(\Gamma_2^T\right)_{\beta\nu}\left(C\tensor{\sigma}^{\delta\lambda}\right)_{\nu\mu}\right]
\nou\left|T_1\right|^2P_\lambda'P_\kappa\bar N\left(P'\right)\gamma_\delta\Gamma_3\gamma_\eta N\left(P\right)
\no
\text{Tr}\left[\Gamma_1\tensor{\sigma}^{\eta\kappa}C\Gamma_2^TC\tensor{\sigma}^{\delta\lambda}\right]
\nou\left|T_1\right|^2P_\lambda'P_\kappa\bar N\left(P'\right)\gamma_\delta\Gamma_3\gamma_\eta N\left(P\right)
\no
\text{Tr}\left[\Gamma_1\tensor{\sigma}^{\eta\kappa}\overset{\leftrightarrow}{\Gamma}_2\tensor{\sigma}^{\delta\lambda}\right]\,;
\end{eqnarray}
\begin{eqnarray}
&&
	\left\langle P'\right|\bar u_\mu\bar u_\nu \bar d_\rho\left|0\right\rangle_A\left\langle0\right|K_H^{\mu\nu\rho\alpha\beta\gamma}\left|0\right\rangle\left\langle 0\right|u_\alpha u_\beta d_\gamma\left|P\right\rangle_V
	\no
	+\left\langle P'\right|\bar u_\mu\bar u_\nu \bar d_\rho\left|0\right\rangle_V\left\langle0\right|K_H^{\mu\nu\rho\alpha\beta\gamma}\left|0\right\rangle\left\langle 0\right|u_\alpha u_\beta d_\gamma\left|P\right\rangle_A
	\nou
	\Big[A_1^*\left(C\gamma_5\slashed{P}'\right)_{\mu\nu}\left(\bar N\left(P'\right)\right)_\rho
	V_1\left(\slashed{P}C\right)_{\alpha\beta}\left(N\left(P\right)\gamma_5\right)_\gamma \Big.
	\no
	\Big.-V_1^*\left(C\slashed{P}'\right)_{\mu\nu}\!\left(\bar N\left(P'\right)\gamma_5\right)_\rho
	A_1\!\!\left(\slashed{P}\gamma_5C\right)_{\alpha\beta}N\left(P\right)_\gamma\! \Big]
	\no
	\Gamma_1^{\mu\alpha}
	\Gamma_2^{\nu\beta}
	\Gamma_3^{\rho\gamma}
	\nou
	-\left(V_1A_1^*+V_1^*A_1\right)\bar N\left(P'\right)\Gamma_3\gamma_5N\left(P\right)\text{Tr}\left[\Gamma_1\slashed{P}\overset{\leftrightarrow}{\Gamma}_2\slashed{P}'\right]\,.
	\no
\end{eqnarray}
\section{Contributing diagrams in terms of the non-perturbative parameters}
\label{appendix:c}
Here we write the integrals used for the calculation of the nucleon form factors $G_M^{p,n}$. The Chernyak-Zhitnitsky formula is 
\begin{eqnarray}
q^4G_M^{p,n}\left(q^2\right)\rightarrow\frac{\left(4\pi\bar\alpha_s\right)^2}{54}\left|f_N\right|^2 I^{p,n}\,,
\end{eqnarray}
where the integrals $I^{p,n}$ up to the second degree polynomials are
\begin{eqnarray}
I^p 
&=&
1400B_0B_1+\frac{2000}{9}B_1^2+1800B_0B_2+\frac{2800}{3}B_1B_2
\no
+1200B_2^2+6600B_0B_3+\frac{22000}{9}B_1B_3+4800B_2B_3
\no
+\frac{18800}{3}B_3^2-\frac{1000}{3}B_0B_4-\frac{2600}{27}B_1B_4-\frac{2200}{9}B_2B_4
\no
-\frac{4600}{9}B_3B_5+\frac{2600}{243}B_4^2-1400B_0B_5-\frac{3500}{9}B_1B_5
\no
-\frac{1100}{3}B_2B_5-\frac{5900}{3}B_3B_5+\frac{4100}{81}B_4B_5+\frac{7700}{27}B_5^2\,;
\n
I^n
&=&
1800B_0^2-1400B_0B_1+\frac{2000}{9}B_1^2-1800B_0B_2
\no
-\frac{2800}{3}B_1B_2-200B_2^2
-1000B_0B_3-\frac{22000}{9}B_1B_3
\no
-2000B_2B_3-\frac{17000}{9}B_3^2+\frac{1000}{3}B_0B_4-\frac{2600}{27}B_1B_4
\no
+\frac{2200}{9}B_2B_4+\frac{4600}{9}B_3B_5+\frac{2600}{243}B_4^2+\frac{2000}{3}B_0B_5
\no
+\frac{3500}{9}B_1B_5+\frac{500}{9}B_2B_5+\frac{6500}{9}B_3B_5-\frac{4100}{81}B_4B_5
\no
-\frac{7250}{81}B_5^2\,,
\end{eqnarray}
where the coefficients $\left\{B_n\right\}_{n=0}^{5}$ are the ones appearing in the expansion of the LCDA $\varphi_N\left(\mathbf{x},Q^2\right)$, and their operative form is given in Eq.~\eqref{eq:parameters}.


\nocite{*}

\bibliography{MicroFFs-ETG-SP-FR-OS-bib}

\end{document}